\documentstyle[12pt]{article}   
\begin{document}
\begin{center}   
\Large
{\bf{Theory of adiabatic fluctuations : third-order noise}}
\end{center}

\begin{center}
{\bf{Suman Kumar Banik, Jyotipratim Ray Chaudhuri and Deb Shankar Ray}}

{\bf{Indian Association for the Cultivation of Science}}

{\bf{Jadavpur , Calcutta 700 032 , INDIA.}}

\end{center}

\vspace{0.5cm}

\begin{abstract}
We consider the response of a dynamical system driven by external
adiabatic fluctuations. Based on the `adiabatic following approximation'
we have made a systematic separation of time-scales to carry out an
expansion in $\alpha |\mu|^{-1}$, where $\alpha$ is the strength of 
fluctuations and $|\mu|$ is the damping rate. We show that probability 
distribution functions obey the differential equations of motion which
contain third order terms ( beyond the usual Fokker-Planck terms ) leading to
non-Gaussian noise. The problem of adiabatic fluctuations in velocity space
which is the counterpart of Brownian motion for fast fluctuations, has been
solved exactly. The characteristic function and the associated probability
distribution function are shown to be of stable form. The linear dissipation
leads to a steady state which is stable and the variances and higher moments
are shown to be finite.
\end{abstract}

\vspace{0.5cm}

{\bf{PACS NO.}} :  05.20 Dd

\newpage


\begin{center}
\large{
{\bf{I. \hspace{0.2cm}Introduction}}}
\end{center}

\vspace{0.5cm}  

In this paper we have discussed the stochastic dynamics of a system driven by
external, adiabatic fluctuations. The opposite counterpart of these processes
correspond to stochastic processes with fast fluctuations which are more 
frequently encountered in physical and chemical sciences.
The classic and celebrated problem of the latter kind is the century-old 
paradigm of Brownian motion first correctly formulated by Einstein [1,2].
In dealing with fast stochastic processes one essentially examines the average 
motion of the system subjected to fast fluctuations ( which may be of external 
or of internal type ) with the following separation of time-scales in mind. If
$\tau_{c}$ is the correlation time of fluctuations which is the shortest 
timescale in the dynamics, compared to coarse-grained timescale $\Delta t$ 
over which one follows the average evolution, then
\begin{eqnarray*}
\tau_{c} \ll \Delta t \ll \frac{1}{|\mu|} \hspace{0.2cm}, \hspace{4.0cm} ({\rm I})
\end{eqnarray*}

\noindent
where $|\mu|^{-1}$ refers to the inverse of the damping rate ( or inverse of 
the largest eigenvalue of the ``unperturbed'' system ). Herein we analyze the average 
dynamics of a general multivariate nonlinear system subjected to {\it external,
adiabatically slow} fluctuations. We have derived the equation of 
motion for evolution of the probability distribution function 
in phase space on a coarse-grained timescale
$\Delta t$ assuming that $\Delta t$ satisfies the following inequality
\begin{eqnarray*}
\frac{1}{|\mu|} \ll \Delta t \ll \tau_{c} \hspace{0.2cm}. \hspace{4.0cm} ( {\rm II})
\end{eqnarray*}

The slow fluctuations characterized by very long correlation time have received
a lot of attention of various workers over the years [2,3]. 
While the treatment
of stochastic differential equations with fast fluctuation is based on the
assumption that there is a very short correlation time such that one is
allowed to make an appropriate expansion in $\alpha \tau_{c}$, where
$\alpha$ is the strength of fluctuation, simplified assumption for dealing with
long correlation time is rather relatively scarce. In general, the problem of 
long correlation time is handled at the expense of severe restriction on the
type of stochastic behavior. For instance, several authors [2,3] have tried
the linear and nonlinear models within the framework of simple Markov processes
of the type, dichotomic process, two-state Markov process, random telegraphic
process, etc. Our strategy here is to follow a {\it{perturbative approach}},
pertaining to the separation of the timescale (II) without {\it{keeping any
above-mentioned restriction on the type of stochastic behavior}}. Based on 
the `adiabatic following approximation' [4] we have recently [5] carried out an 
expansion in $\alpha |\mu|^{-1}$ to obtain a linear differential equation 
for the average solution. In this paper we extend this analysis to treat
{\it{nonlinear}} stochastic differential equations for construction of
appropriate master equations. The perturbative expansion
is essentially a counterpart of expansion in $\alpha \tau_{c}$ as dealt
in the case of fast fluctuations [2]. The difference 
between the two expansion schemes
lies in identification of two distinct shortest time-scales in the dynamics of
the two cases. In the case of fast fluctuations it is $\tau_{c}$, whereas the
corresponding role is played by $|\mu|^{-1}$ in adiabatic fluctuations.

We have shown that the equation of motion in phase space for probability distribution
function contains beyond the ordinary Fokker-Planck terms, third order
derivative terms. As shown by Pawula [6] for one dimensional case, an equation 
with third order derivative terms is in contradiction to the positivity for
transition probability for short time. However it is well known that finite
derivative terms of order larger than two may be quite useful in different
occasions [7-8], e.g., in the treatment of optical bistability described in terms
of quasi-distribution function of Wigner in quantum optics [7] or explaining
trimolecular reactions using Poisson representation of Fokker-Planck equation,
also in one dimensional random walk with boundary within a scheme of expansion 
of master equation.
Although at this stage of development a clear general probabilistic interpretation
in terms of any real stochastic process is lacking [7] one can 
identify the noise terms with distinct characteristics for such processes. In 
a similar spirit we are led to the conclusion in the present context that 
adiabatic fluctuations
give rise to third order non-Gaussian noise terms beyond the usual Fokker-Planck
terms.

The central result of this paper is the solution of the problem of adiabatic
fluctuations in velocity space, which is the counterpart of Brownian dynamics
for rapid fluctuations. We have shown that the characteristic function obeys a 
simple third order differential equation. This can be solved exactly to obtain 
a probability distribution of stable form which for small arguments displays a 
power law behavior. It is also important to note that the linear dissipation leads to a
stable steady state distribution. However, the fluctuation being external
the energy supplied by this cannot be balanced by dissipation and as such 
there is no fluctuation-dissipation relation in this case. Furthermore, the
non-Gaussian statistical characteristics can be obtained from the calculation 
of variances and higher moments which are shown to be finite. We thus conclude 
that although in many cases third order noise makes the probabilistic 
consideration truly difficult, the systems driven by adiabatic fluctuations
display a distinct non-Gaussian stochastic behavior is
amenable to understanding in simple probabilistic terms.
Occasionally wherever possible we allow ourselves a fair comparison with Levy
processes [9,11] and point out the essential differences.

The outlay of the paper is as follows : In the next section we review the
basic aspects of adiabatic fluctuations in linear processes as dealt in our
earlier paper [5]. The two basic
assumptions, e.g., the adiabatic following approximation and the decoupling 
approximation as well as validity and convergence of perturbative expansion 
were discussed in detail
in the earlier paper [5]. To make this paper self-contained and readable we 
review its salient features. In
Sec.III we extend the treatment to nonlinear equations. The equations in 
phase space have been derived in Sec.IV. The counterpart of Brownian motion
in velocity space for slow fluctuations have been treated in Sec.V. Explicit
solution for the probability distribution function and the approach to
equilibrium have been discussed. The paper is concluded in Sec.VI.

\newpage


\begin{center}
\large{
{\bf{II. \hspace{0.2cm} Linear processes with adiabatic fluctuations}}}
\end{center}

\vspace{0.5cm}

To begin with we have considered the following linear equation,
\begin{equation}
{\dot{u}}=\{ {\bf{A}}_{0} + \alpha {\bf{A}}_{1}(t)\} u
\hspace{0.05cm},
\end{equation}

\noindent
where $u$ is a vector with $n$ components, ${\bf{A}}_{0}$ is a constant 
matrix of dimension $n\times n$ with negative real eigenvalues and 
${\bf{A}}_{1}(t)$ is a random matrix, 
$\alpha$ is a parameter which measures the strength of fluctuation.

It is convenient to assume that ${\bf{A}}_{1}(t)$ is a stationary process 
with $\langle {\bf{A}}_{1}(t)\rangle=0$. Eq.(1) sets the two  
time scales of the system, measured by the inverse of the largest eigenvalue
of the matrix ${\bf{A}}_{0}$  and the time scale of fluctuations 
of ${\bf{A}}_{1}(t)$ (more precisely correlation time of fluctuation). 
In the problem of Brownian motion where one deals with very fast fluctuations 
such that correlation time $\tau_{c}$ is essentially the shortest time scale
in the dynamics, one thus follows the evolution of the average 
$\langle u \rangle$ on a coarse-grained timescale.

Before proceeding further we now make two remarks at this stage : 
First, since in the present context we have considered a stochastic process in which the 
fluctuations are weak and adiabatically slow, ${\bf{A}}_{1}(t)$ is  an 
adiabatic stochastic process. Therefore the usual procedure of systematic 
expansion in $\alpha \tau_{c}$ which relies on smallness of $\tau_{c}$, is 
not valid. We thus take resort to a different approach based on an expansion 
in $\alpha |\mu|^{-1}$, where $|\mu|$ refers to the largest eigenvalue
of ${\bf A}_{0}$ matrix. Second, we {\it do not} make any 
{\it a priori assumption
about the nature of the stochastic process}, such as, Gaussian or dichotomic
etc. The only assumption that have been made about the stochastic process is 
that the inverse of the damping rate is much shorter compared to the correlation
time of fluctuations ${\bf A}_{1}(t)$.

As a first step we introduce an interaction representation as given
by,
\begin{equation}
u(t)=\exp({\bf{A}}_{0} t) v(t) \hspace{0.2cm},
\end{equation}

\noindent
and applying it to Eq.(1) we obtain,
\begin{equation}
{\dot{v}}=\alpha {\bf{V}}(t) v \hspace{0.05cm},
\end{equation}

\noindent
where, 
\begin{equation}
{\bf{V}}(t)=\exp(-{\bf{A}}_{0} t) {\bf{A}}_{1}(t)\exp({\bf{A}}_{0} t) \hspace{0.05cm}.
\end{equation}

On integration Eq.(3) yields,
\begin{equation}
v(t)=v(0)+\alpha\int_{0}^{t} {\bf{V}}(t')v(t') dt' \hspace{0.05cm} .
\end{equation}

On iterating Eq.(5) once, we are led to an ensemble average equation 
of the form,
\begin{equation} 
\langle v(t)\rangle=v(0)+\alpha^{2}\int_{0}^{t} dt'\int_{0} 
^{t'} d t'' \langle {\bf{V}}(t'){\bf{V}}(t'') v(t'')\rangle \hspace{0.05cm} .
\end{equation} 

The equation is still exact since no second order approximation has been used.

Now taking the time derivative of $v(t)$ we arrive at the following
integrodifferential equation in which the initial value $v(0)$ no
longer appears,
\begin{equation} 
\frac{d}{d t}\langle v(t)\rangle=\alpha^{2}\int_{0}^{t}  
\langle {\bf{V}}(t){\bf{V}}(t') v(t')\rangle dt' \hspace{0.05cm}.
\end{equation} 

Making use of a change of integration variable $t'=t-\tau$ and 
reverting back to the original representation we obtain
\begin{equation} 
\frac{d}{d t}\langle u(t)\rangle={\bf{A}}_{0}\langle u\rangle
+\alpha^{2}\int_{0}^{t}  
\langle {\bf{A}}_{1}(t)\exp({\bf{A}}_{0}\tau){\bf{A}}_{1}(t-\tau)
u(t-\tau) \rangle d\tau \hspace{0.05cm} .
\end{equation} 

The {\it adiabatic following assumption} (see the discussion at the end of 
this section), that ${\bf{A}}_{1}(t)$ and the components
of $u(t)$ vary slowly on the scale of inverse of ${\bf{A}}_{0}$, can 
now be utilized. Following Crisp [4] we note that a Taylor series expansion of
${\bf{A}}_{1}(t-\tau) u(t-\tau)$ in the average $\langle\ldots\rangle$
of the $\alpha^{2}$-term in Eq.(8) allows us to reduce the above equation
to the following form,
\begin{equation} 
\frac{d}{d t}\langle u(t)\rangle={\bf{A}}_{0}\langle u\rangle
+\alpha^{2}\sum_{n=o}^{\infty}\frac{(-1)^{n}}{n !}  
\langle {\bf{A}}_{1}(t) {\bf I}_{n} 
\frac{d^{n}}{d t^{n}}[{\bf{A}}_{1}(t) u(t)] 
\rangle \hspace{0.5cm} . 
\end{equation} 

\noindent
${\bf I}_n$ can also be written as
\begin{eqnarray*}
I_n^{ik} & = & \int_0^\infty d\tau \; \tau^n \sum_j D_{ij} \; 
e^{\mu_{jj}\tau} \; D_{jk}^{-1} \nonumber\\
& = & \sum_j D_{ij} \; \frac{n!}{\mu_{jj}^{n+1}} \; D_{jk}^{-1} \; ,  
\; \; \; \; \; {\rm Re}\;  \mu_{jj} < 0 \; \; .
\end{eqnarray*}

Eq.(9) can then be rewritten in the form
\begin{equation} 
\frac{d}{d t}\langle u(t)\rangle={\bf{A}}_{0}\langle u\rangle
+\alpha^{2}\sum_{n=o}^{\infty}(-1)^{n}  
\langle {\bf{A}}_{1}(t) {\bf{D}} {\bf{E}}_{n+1}{\bf{D}}^{-1}
\frac{d^{n}}{d t^{n}}[ {\bf{A}}_{1}(t) u(t) ] \rangle \hspace{0.05cm},
\end{equation} 

\noindent
where we use
\begin{equation}
{\bf I}_n=n!\hspace{0.1cm}{\bf{D}}\hspace{0.1cm}{\bf{E}}_{n+1}
\hspace{0.1cm}{\bf{D}}^{-1} \hspace{0.05cm}.
\end{equation}

Here ${\bf{D}}$ is a matrix which diagonalises ${\bf{A}}_{0}$ and
\begin{eqnarray*}
{\bf{E}}_{n+1}=\left(\begin{array}{ccc}
\frac{1}{\mu_{11}^{n+1}} & & 0\\
& \ddots &\\
0 & & \frac{1}{\mu_{jj}^{n+1}}\end{array}\right ) 
\end{eqnarray*}

\noindent
and $\mu_{jj}$ are the eigenvalues of ${\bf{A}}_{0}$.

Although the Eq.(10) involves an infinite series it is expected to yield
useful result in the adiabatic following approximation. Under this 
approximation the quantity $[{\bf{A}}_{1}(t) u(t)]$ varies very little
(such that $\frac{d^n}{dt^n} ({\bf A}_{1} u)$ in Eq.(10) is small)
and also since $|\mu_{jj}|$ in ${\bf E}_{n+1}$ is large the series in Eq.(10)
( which is thus an expansion in $\alpha |\mu|^{-1}$ ) converges rapidly.
Keeping only the two lowest order terms we arrive at,
\begin{eqnarray} 
\frac{d}{d t}\langle u(t)\rangle={\bf{A}}_{0}\langle u\rangle
+\alpha^{2}\langle {\bf{A}}_{1}(t){\bf{X}}_{1}{\bf{A}}_{1}(t) u(t)
\rangle -\alpha^{2}\langle {\bf{A}}_{1}(t){\bf{X}}_{2}{\dot{\bf{A}}_{1}}
(t) u(t)\rangle\nonumber\\
-\alpha^{2}\langle {\bf{A}}_{1}(t){\bf{X}}_{2}{\bf{A}}_{1}(t){\dot{u}}(t)
\rangle
\end{eqnarray} 

\noindent
where,
\begin{eqnarray*}
{\bf{X}}_{n+1}={\bf{D}}\hspace{0.1cm}{\bf{E}}_{n+1}\hspace{0.1cm}
{\bf{D}}^{-1} \hspace{0.05cm}.
\end{eqnarray*}

It is evident that the average $\langle {\dot{u}} \rangle$ is related to 
a more complicated average. Following Bourret [10], we now implement the decoupling 
approximation. This allows us to break up the average as a product of averages.
Keeping terms only of the order of $\alpha^2$ we obtain

\newpage

\begin{eqnarray} 
\frac{d}{dt}\langle u(t)\rangle=\left \{ {\bf{A}}_{0}
+\alpha^{2}[ \langle {\bf{A}}_{1}(t){\bf{X}}_{1}{\bf{A}}_{1}(t)\rangle-
\langle {\bf{A}}_{1}(t){\bf{X}}_{2}{\dot{\bf{A}}_{1}}(t)\rangle
\right. \nonumber\\
\left.-\langle {\bf{A}}_{1}(t){\bf{X}}_{2}{\bf{A}}_{1}(t)\rangle{\bf{A}}_{0}]
\right\} \langle u(t)\rangle \hspace{0.05cm} .
\end{eqnarray} 

Thus the average of $u(t)$ obeys a nonstochastic differential equation
in which the effect of weak adiabatic fluctuations is accounted for by
renormalizing ${\bf{A}}_{0}$ through the addition of constant terms
of the order of $\alpha^{2}$. 

The implementation of Bourret's decoupling approximation [10] is a major step 
for almost any treatment of multiplicative noise upto date [2,3,12]. This
is because of the fact that the average of a product of stochastic quantities 
does not factorize into the product of averages, although it has been argued 
that good approximations can be derived by assuming such factorization.
In the case of fast fluctuations it has been justified if the driving stochastic
noise has a short correlation time such that Kubo number $\alpha^2\tau_{c}$ is
very small in the cummulant expansion scheme ( an expansion in $\alpha \tau_{c}$
). The factorization has been shown to be exact in the limit of zero 
correlation time and in some cases of specific noise processes [3,12]
and the solution for the average can be found exactly. 

In contrast to cummulant expansion (valid in the case of fast fluctuation
which relies on an expansion in $\alpha \tau_c$) the present scheme of 
adiabatic following approximation results in a perturbation series, where the
n-th term is of order 
$\alpha \frac{d^n}{dt^n} [{\bf A}(t)u(t)] /\mu^{n+1}_{jj}$ and the convergence 
of the series is assured since the numerator varies little in the scale of
$1/|\mu^{n+1}_{jj}|$. Eq.(13) is a result of decoupling approximation
employed in this expansion scheme. If one neglects the free motion due to
${\bf A}_0$ term then Eq.(13), which gives the lowest order evolution, 
asserts that
\begin{eqnarray*}
\frac{d}{dt} \langle u\rangle \sim \frac{\alpha^2}{|\mu|} 
\; \langle u\rangle \; \; .
\end{eqnarray*}

The contribution of $|\mu|^{-1}$ is derived from ${\bf X}_1$ of the first 
term in Eq.(13), (i.e., due to ${\bf E}_{n+1}$ matrix). 
Note that because of full integration over $\tau$ 
in moving from Eq.(8) to Eq.(9) correlation time $\tau_c$ does not appear
in Eq.(13) and the time-scale set by the dynamics is $|\mu|^{-1}$ only. For
a fast process on the other hand the counterpart of the last relation is [12]
\begin{eqnarray*}
\frac{d}{dt}\langle u\rangle \sim \alpha^2 \tau_c \; \langle u\rangle \; \; .
\end{eqnarray*}

It is also easy to calculate the relative error made in the decoupling
approximation. We first note that Eq.(13) is obtained from Eq.(8). To the
second order it means omitting terms of the order 
$(\alpha \Delta t)^3$ and higher (where $\Delta t$ is the coarse-grained
time-scale over which $\langle u\rangle$ evolves). As the lower bound of 
$\Delta t$ is determined by $|\mu|^{-1}$, it implies that we neglect terms of
the order $(\alpha |\mu|^{-1})^3$ in the evolution equation. The relative 
error made in the decoupling approximation is thus of the order 
$(\alpha |\mu|^{-1})^3$ which is well within the order of lowest order
evolution. We thus see that the adiabatic expansion 
is an expansion in $\alpha |\mu|^{-1}$ and the decoupling approximation
in the slow fluctuation is valid where
$\alpha^2 |\mu|^{-1}$ is very small. Thus $u(t)$
in the average ( in the right hand side of Eq.(12) ) is realized as an
average $\langle u(t)\rangle$ ( which varies in the coarse-grained timescale
$\Delta t$ ) in Eq.(13) pertaining to the separation of the time-scales in 
the inequality (II) in Sec.I. 

Before closing this section a few pertinent points regarding the
notion of ``{\it adiabatic following approximation}" and its genesis may
be noted. The notion has acquired special relevance in the quantum optical
context where one is concerned with a two-level atom interacting with
single mode electromagnetic field. The model is described by the standard 
Bloch equations, where the field strength varies slowly on the time-scale 
of inverse of the damping constant or the frequency detuning between
the atom and the field. If the field is varying adiabatically enough, 
then the population inversion of the Bloch vector components would 
{\it follow the field adiabatically} in
going from ground to upper state, i.e., the ground state population is 
adiabatically inverted. The term ``{\it adiabatic following}" is thus used
to describe collectively the associated experimental phenomena [19].

\newpage


\begin{center}
\large{
{\bf III. \hspace{0.2cm} Probabilistic considerations : Extension to nonlinear
equations}}
\end{center}

\vspace{0.5cm}

We now generalize Eq.(1) to a stochastic nonlinear differential equation 
written in the following form
\begin{equation}
\dot{u}_{\nu} = F_{\nu} (\{ u_{\nu} \}, t ; \xi (t) ) \hspace{0.2cm} ,\hspace{0.2cm}
\nu = 1, 2, \ldots ,N \hspace{0.2cm}.
\end{equation}

The above equation determines a stochastic process with some initial given 
condition $\{ u_{\nu}(0)\}$. $\xi (t)$ is the adiabatic stochastic process.
It may be pointed out that the treatment given in the last section cannot be 
extended directly to this equation to obtain an equation for average
$\langle u \rangle$ since nonlinearity in Eq.(14) results in higher moments.
However, it is possible to transform the nonlinear problem to a linear one
if one considers the motion of a representative point-$u$ in n-dimensional
space ($u_{1}\ldots u_{n}$) as governed by Eq.(14). The equation of continuity
, which expresses the conservation of points determines the variation of
density in time,
\begin{equation}
\frac{\partial \rho (u,t)}{\partial t} = -\sum_{\nu} \frac{\partial}{\partial
u_{\nu}} F_{\nu} (\{ u_{\nu} \},t;\xi(t)) \rho (u,t)
\end{equation}

\noindent
or more compactly
\begin{equation}
\frac{\partial \rho}{\partial t} = -\nabla \cdot {\bf F}\rho \hspace{0.2cm}.
\end{equation}

Eq.(15) is a linear stochastic differential equation and is an ideal candidate
for the method discussed in the last section for the linear case. We emphasize
here that the basis of the present analysis is essentially the two
approximations as introduced earlier and {\it{no further approximation}} is
needed to extend the analysis to nonlinear domain.

{\bf F} can now be split as
\begin{equation}
{\bf F} (\{ u_{\nu} \},t;\xi(t)) = {\bf F}_{0} (\{ u_{\nu} \}) 
+ \alpha {\bf F}_{1} (\{ u_{\nu} \},t;\xi(t)) \hspace{0.2cm}, 
\end{equation}

\noindent
where ${\bf F}_{0} (\{ u_{\nu} \})$ is the constant part and  
${\bf F}_{1} (\{ u_{\nu} \},t;\xi(t)) $ is the random part with
$\langle {\bf F}_{1}(t)\rangle =0$; $\alpha$ is the parameter defined earlier
which measures the strength of fluctuation. Eq.(16) therefore takes the 
following form,
\begin{equation}
\dot{\rho}(u,t) = ( {\bf A}_{0} +\alpha {\bf A}_{1} ) \rho(u,t) \hspace{0.2cm},
\end{equation}

\noindent
where ${\bf A}_{0} = -\nabla\cdot {\bf F}_{0}$ and 
$ {\bf A}_{1} =-\nabla \cdot {\bf F}_{1}$. 
The symbol $\nabla$ is used for the operator that
differentiates everything that comes after it with respect to $u$.

With the above identification of ${\bf A}_{0}$ and ${\bf A}_{1}$ we are now in 
a position to apply the fundamental Eq.(9) derived in the earlier section,
to Eq.(18). We have
\begin{equation}
\frac{\partial}{\partial t} P(u,t) = \left [ -\nabla\cdot {\bf F}_{0} P +
\alpha^{2}\sum_{n=0}^{\infty} \frac{(-1)^n}{n !} \langle - \nabla \cdot
{\bf F}_{1} {\bf I}_{n} \frac{d^n}{d t^n} (-\nabla\cdot {\bf F}_{1}P)\rangle
\right] \hspace{0.2cm},
\end{equation}

\noindent
where $\langle \rho(u,t)\rangle = P(u,t)$ and also
\begin{equation}
{\bf I}_{n}=\int_{0}^{\infty} d\tau e^{-\tau\nabla\cdot {\bf F}_{0}} \tau^{n}
\hspace{0.2cm}.
\end{equation}

Adiabatic following approximation may now be invoked again in the spirit of 
earlier treatment in Sec.II to obtain

\newpage

\begin{eqnarray}
\frac{\partial}{\partial t} P(u,t) = & - & \nabla\cdot \left [ {\bf F}_{0}
+  \alpha^{2} \langle {\bf F}_{1} {\bf I}_{0} \nabla\cdot {\bf F}_{1} \rangle 
-  \alpha^{2} \langle {\bf F}_{1} {\bf I}_{1} \nabla\cdot {\dot{\bf F}}_{1} \rangle 
\right. \nonumber\\
& + & \left. \alpha^{2} \langle {\bf F}_{1} {\bf I}_{1} \nabla\cdot {\bf F}_{1} 
\nabla\cdot {\bf F}_{0}\rangle \right ] P(u,t) \hspace{0.2cm},
\end{eqnarray}

\noindent
where we keep terms of the order of $\alpha^2$ for $n=0$ and $1$ of the
series in Eq.(19).

Our next task is to simplify further the expressions for the averages in Eq.(21).
To this end we first note that the operator $\exp(-\tau\nabla\cdot {\bf F}_{0})$
provides the solution of the equation
\begin{equation}
\frac{\partial f(u,t)}{\partial t} = -\nabla\cdot {\bf F}_{0} f(u,t) \hspace{0.2cm},
\end{equation}

\noindent
($f$ signifies the unperturbed part of $P$) which can be found explicitly in
terms of characteristic curves. The equation
\begin{eqnarray*}
\dot{u}={\bf F}_{0} (u)
\end{eqnarray*}

\noindent
for fixed $t$ determines a mapping from $u(\tau=0)$ to $u(\tau)$, i.e.,
$u \rightarrow u^{\tau}$ with inverse $(u^{\tau})^{-\tau}=u$. The solution of 
Eq.(22) is
\begin{equation}
f(u,t)=f(u^{-t},0) \left | \frac{d (u^{-t})}{d(u)} \right | = e^{-t\nabla\cdot
{\bf F}_{0}} f(u,0) \hspace{0.2cm},
\end{equation}

\noindent
$\left | \frac{d (u^{-t})}{d(u)} \right |$ being a Jacobian determinant. The
effect of $\exp(-t\nabla\cdot {\bf F}_{0})$ or $f(u)$ is as follows
\begin{equation}
\exp(-t\nabla\cdot {\bf F}_{0}) f(u,0) = f(u^{-t},0) 
\left | \frac{d (u^{-t})}{d(u)} \right | \hspace{0.2cm}.
\end{equation}

The relation (24) may be used to simplify the average in Eq.(21). We thus have
\begin{eqnarray}
\langle \nabla\cdot {\bf F}_{1} {\bf I}_{0} \nabla\cdot {\bf F}_{1} \rangle =
\nabla\cdot \int_{0}^{\infty} \langle {\bf F}_{1}\nabla_{-\tau}\cdot
{\bf F}_{1}(u^{-\tau})\rangle \left | \frac{d u^{-\tau}}{d u} \right | 
d \tau \hspace{0.2cm},\\
\nonumber\\
\langle \nabla\cdot {\bf F}_{1} {\bf I}_{1} \nabla\cdot {\dot{\bf F}}_{1} \rangle =
\nabla\cdot \int_{0}^{\infty} \tau \langle {\bf F}_{1}\nabla_{-\tau}\cdot
{\dot{\bf F}}_{1}(u^{-\tau})\rangle \left | \frac{d u^{-\tau}}{d u} \right | 
d \tau \hspace{0.2cm},\\
\nonumber\\
\langle \nabla\cdot {\bf F}_{1} {\bf I}_{1} \nabla\cdot {\bf F}_{1} 
\nabla \cdot {\bf F}_{0} \rangle =
\nabla\cdot \int_{0}^{\infty} \tau \langle {\bf F}_{1}\nabla_{-\tau}\cdot
{\bf F}_{1}(u^{-\tau}) \nabla_{-\tau}\cdot {\bf F}_{0} (u^{-\tau}) \rangle
\left | \frac{d u^{-\tau}}{d u} \right | d \tau \hspace{0.2cm}.
\end{eqnarray}

The use of Eq.(25)-(27) reduces Eq.(21) to a more tractable form,
\begin{eqnarray}
\frac{\partial}{\partial t} P(u,t) = & - & \nabla\cdot \left \{ {\bf F}_{0}  
- \alpha^{2}\int_{0}^{\infty} \langle {\bf F}_{1}\nabla_{-\tau}\cdot
{\bf F}_{1}(u^{-\tau})\rangle \left | \frac{d u^{-\tau}}{d u} \right | 
d \tau  \right. \nonumber \\
\nonumber\\
& + & \alpha^{2} \int_{0}^{\infty} \tau \langle {\bf F}_{1}\nabla_{-\tau}\cdot
{\dot{\bf F}}_{1}(u^{-\tau})\rangle \left | \frac{d u^{-\tau}}{d u} \right | 
d \tau \nonumber\\
\nonumber\\
& - & \left. \alpha^{2}\int_{0}^{\infty} \tau \langle {\bf F}_{1}\nabla_{-\tau}\cdot
{\bf F}_{1}(u^{-\tau}) \nabla_{-\tau}\cdot {\bf F}_{0} (u^{-\tau}) \rangle
\left | \frac{d u^{-\tau}}{d u} \right | d \tau 
\right \} P(u,t) \hspace{0.04cm},
\end{eqnarray}

\noindent
where $\nabla_{-\tau}$ denotes the differential with respect to $u_{-\tau}$.
Eq.(28) is our basic result in this section. The equation is second order in 
$\alpha$, i.e., of the order of $\alpha^{2} |\mu|^{-1}$, where $|\mu|$ refers to the
eigenvalue of ${\bf A}_{0}$.  In our earlier communication [5] we have shown
the convergence of the series in $\alpha |\mu|^{-1}$, pertaining to the 
separation of the time-scales implied in II in Sec.I. We also remark that
it is possible to extend the treatment to higher order, in general.
It is also to be noted that the equation involves 
three differentiation of $P(u,t)$ with respect to the components of $u$ and is a
third order equation. The appearance of third-order noise beyond the usual 
Fokker-Planck terms is a characteristic of the process we consider here. We
discuss this aspect in more detail in the following two sections.

\vspace{0.5cm}


\begin{center}
\large{
\bf{IV.\hspace{0.2cm}Adiabatic stochasticity in phase space}}
\end{center}

\vspace{0.5cm}

We now consider the motion of a particle in one dimension subjected to a force
$K(x)$ depending on the position $x$, a frictional force $-\beta \dot{x}$
and a stochastic force $\alpha \xi(t)$. Here $\beta$ is a measure of damping
of the system and $\alpha$ is the strength of adiabatically slow
fluctuations $\xi(t)$. We thus write
\begin{equation}
m\ddot{x} +\beta\dot{x} = K(x) +\alpha\xi(t)\hspace{0.2cm}.
\end{equation}

The corresponding problem of fast fluctuation $\alpha\xi(t)$ was studied
by Kramers [13] as a model of simple chemical reactions and by 
Bixon and Zwanzig [14] as a model for fluctuating nonlinear systems.

For simplicity we set $m=1$ and $\dot{x}=v$. Then the two components of $u$
in this example are $x$ and $v$. Taking Eq.(17) into account we have
\begin{equation}
\left. \begin{array}{ll}
F_{0x}=v & F_{1x}=0 \\
F_{0v}=-\beta v +K(x) & F_{1v}=\alpha\xi(t)
\end{array} \right \} \hspace{0.2cm}.
\end{equation}

By considering a small variation of $v$ in time $\tau$, one obtain (from
the unperturbed version of Eq.(29)) the Jacobian determinant for the
``unperturbed'' mapping $u\rightarrow u^{\tau}$
\begin{equation}
\left | \frac{du^{-\tau}}{du} \right | \equiv 
\left | \frac{d (x^{-\tau},v^{-\tau})}{d(x,v)} \right | = 1+\beta\tau +
{\cal O}(\tau^2)
\end{equation}

\noindent
and
\begin{equation}
\left. \begin{array}{ccc}
\frac{\partial}{\partial v^{-\tau}} & = & (1-\beta\tau)\frac{\partial}
{\partial v}+\tau \frac{\partial}{\partial x}+{\cal O}(\tau^2) \\
\frac{\partial}{\partial x^{-\tau}} & = & \frac{\partial}{\partial x}
+\tau \frac{\partial K(x)}{\partial x}\frac{\partial}{\partial v}
+{\cal O}(\tau^2) 
\end{array} \right \}\hspace{0.2cm}.
\end{equation}

The Eq.(28) now reduces to the following form
\begin{eqnarray}
\frac{\partial}{\partial t} P(x,v,t) = - \nabla\cdot\left \{ {\bf F}_{0}
-\alpha^2 \int_{0}^{\infty} \langle {\bf F}_{1}\nabla_{-\tau}\cdot {\bf F}_{1}
(x^{-\tau},v^{-\tau}) \rangle
\left | \frac{d (x^{-\tau},v^{-\tau})}{d(x,v)} \right | d\tau \right.\nonumber\\
\nonumber\\
+ \alpha^2 \int_{0}^{\infty} \tau \langle {\bf F}_{1}\nabla_{-\tau}\cdot 
{\dot{\bf F}}_{1} (x^{-\tau},v^{-\tau}) \rangle
\left | \frac{d (x^{-\tau},v^{-\tau})}{d(x,v)} \right | d\tau \nonumber\\
\nonumber\\
- \left. \alpha^2 \int_{0}^{\infty} \tau \langle {\bf F}_{1}\nabla_{-\tau}\cdot 
{\bf F}_{1} (x^{-\tau},v^{-\tau}) \nabla_{-\tau}\cdot {\bf F}_{0}
(x^{-\tau},v^{-\tau}) \rangle
\left | \frac{d (x^{-\tau},v^{-\tau})}{d(x,v)} \right | d\tau \right \}
P(x,v,t).
\end{eqnarray}

Making use of relations (30-32) one may reduce the terms on the right hand 
side of Eq.(33) to more simplified forms. Thus
\begin{equation}
-\nabla\cdot {\bf F}_{0} P(x,v,t)= -v\frac{\partial P}{\partial x} + \beta
\frac{\partial}{\partial v}(vP)-K(x)\frac{\partial P}{\partial v}
\hspace{0.2cm},
\end{equation}

\begin{equation}
\alpha^2 \nabla\cdot \int_{0}^{\infty} \langle {\bf F}_{1}\nabla_{-\tau}\cdot 
{\bf F}_{1} (x^{-\tau},v^{-\tau}) \rangle
\left | \frac{d (x^{-\tau},v^{-\tau})}{d(x,v)} \right | d\tau  P(x,v,t) =
\alpha^{2} {\tilde c}_{0}\frac{\partial^2 P}{\partial v^2} +
\alpha^{2}{\tilde c}_{1} \frac{\partial^2 P}{\partial v \partial x} 
\hspace{0.2cm},
\end{equation}

\begin{equation}
\alpha^2 \nabla\cdot\int_{0}^{\infty} \tau \langle {\bf F}_{1}\nabla_{-\tau}
\cdot {\dot{\bf F}}_{1} (x^{-\tau},v^{-\tau}) \rangle
\left | \frac{d (x^{-\tau},v^{-\tau})}{d(x,v)} \right | d\tau P(x,v,t) =
-\alpha^2 {\tilde c}_{2} \frac{\partial^2 P}{\partial v^2}
\hspace{0.2cm},
\end{equation}

\newpage

\begin{eqnarray}
\alpha^2 \nabla\cdot \int_{0}^{\infty} \tau \langle {\bf F}_{1}
\nabla_{-\tau}\cdot 
{\bf F}_{1} (x^{-\tau},v^{-\tau}) \nabla_{-\tau}\cdot {\bf F}_{0}
(x^{-\tau},v^{-\tau}) \rangle
\left | \frac{d (x^{-\tau},v^{-\tau})}{d(x,v)} \right | d\tau 
P(x,v,t) \nonumber\\
\nonumber\\
=\alpha^2 {\tilde c}_{1} \left [ 2 \frac{\partial^2 P}{\partial v \partial x}
+ v \frac{\partial^3 P}{\partial v^2 \partial x} + K(x) \frac{\partial^3 P}
{\partial v^3} - \beta \frac{\partial^3}{\partial v^3}(vP) \right ]
\hspace{0.2cm},
\end{eqnarray}

\noindent
where
\begin{equation}
\left. \begin{array}{ccc}
{\tilde c}_{0} & = & \int_{0}^{\infty} \langle\xi (t)\xi (t-\tau )\rangle d\tau \\
{\tilde c}_{1} & = & \int_{0}^{\infty} \tau \langle\xi (t)\xi (t-\tau )\rangle d\tau \\
{\tilde c}_{2} & = & \int_{0}^{\infty} \tau \langle\xi (t) {\dot{\xi}} (t-\tau )\rangle d\tau
\end{array} \right \} \hspace{0.2cm}.
\end{equation}

The final equation for the average motion corresponding to an adiabatic
stochastic evolution in phase space is,
\begin{eqnarray}
\frac{\partial}{\partial t} P(x,v,t) =
& - & v\frac{\partial P}{\partial x} + \beta
\frac{\partial}{\partial v}(vP)-K(x)\frac{\partial P}{\partial v}
+ \alpha^{2} ({\tilde c}_{0}-{\tilde c}_{2}) \frac{\partial^2 P}{\partial v^2} 
+ 3 \alpha^{2} {\tilde c}_{1} \frac{\partial^2 P}{\partial v \partial x} \nonumber\\
\nonumber\\
& + & \alpha^2 {\tilde c}_{1} \left [
v \frac{\partial^3 P}{\partial v^2 \partial x} + K(x) \frac{\partial^3 P}
{\partial v^3} - \beta \frac{\partial^3}{\partial v^3}(vP) \right ]
\hspace{0.2cm}.
\end{eqnarray}

The remarkable departure from the standard form of Fokker-Planck equation
is thus evident in Eq.(39) since it contains third derivative terms.
The magnitude of their contribution is dependent on how much `unperturbed'
$x$ and $v$ vary during $\tau$ which is of the order of $|\mu|^{-1}$.
We also point out that in the above derivation Bourret's decoupling
approximation [10] has been used as in the treatment of linear equation in
Sec.II.

\vspace{0.5cm}


\begin{center}
\large{
\bf{V.\hspace{0.2cm}Adiabatic fluctuations in velocity space}}
\end{center}

\vspace{0.5cm}

We now consider the motion of a particle with velocity $v$ in presence of
fluctuations $\alpha \xi (t)$ which is adiabatically slow. The 
equation of motion is given by
\begin{equation}
\dot{v} = -\beta v + \alpha \xi (t) \hspace{0.2cm}.
\end{equation}

The corresponding problem of a Brownian particle with fast fluctuations is a
century-old problem in physical science, in general. Following the procedure
described in the earlier section we first identify the perturbed and the
unperturbed part of ${\bf F}$, i.e.,
\begin{equation}
\begin{array}{ccc}
{\bf F}_{0}= -\beta v & , & {\bf F}_{1}=\alpha\xi (t)
\end{array}
\end{equation}

\noindent
and calculate the Jacobian $\left |\frac{d v^{-\tau}}{d v}\right |$ for the
mapping $v\rightarrow v^{\tau}$ for the `unperturbed' motion
\begin{equation}
\left |\frac{d v^{-\tau}}{d v}\right | = e^{\beta\tau} \hspace{0.2cm}.
\end{equation}

Also note that
\begin{equation}
\nabla_{-\tau} = e^{-\beta\tau} \frac{\partial}{\partial v} \hspace{0.2cm}.
\end{equation}

The evolution of the probability distribution function
$P(v,t)$ is then given by (terms of the order $\alpha^2$)
\begin{equation}
\frac{\partial}{\partial t} P(v,t) = \beta \frac{\partial}{\partial v} (vP)
+\alpha^2 c_{12} \frac{\partial^2 P}{\partial v^2} - \alpha^2 c_{3}
\frac{\partial^3}{\partial v^3} (vP)\hspace{0.2cm},
\end{equation}

\noindent
where
\begin{equation}
\left. \begin{array}{ccc}
c_{12} & = & c_{1}-c_{2}\\
c_{1} & = & \int_{0}^{\infty} \langle \xi (t) \xi (t-\tau) \rangle d\tau\\
c_{2} & = & \int_{0}^{\infty} \tau \langle \xi (t) {\dot{\xi}} (t-\tau) \rangle d\tau\\
c_{3} & = & \int_{0}^{\infty} \tau \langle \xi (t) \xi (t-\tau) \rangle d\tau
\end{array} \right \} \hspace{0.2cm}.
\end{equation}

While in the absence of the third term, the first two terms on the right hand 
side of Eq.(44) correspond to drift and diffusion terms in the Fokker-Planck 
description of an Ornstein-Uhlenbeck process, the third derivative term 
precludes the possibility of any straight-forward interpretation of the 
equation. Similar equations with third order noise, although not very common, 
however may be encountered [7] in the treatment of trimolecular reactions and 
also in quantum optics describing optical bistability in terms of associated 
Wigner distribution function for the reduced density operator in 
symmetrical ordering for the radiation field. 

We now return to the Eq.(44) which after some modification becomes
\begin{equation}
\frac{\partial}{\partial t} P(v,t) = \beta \frac{\partial}{\partial v}
[vP(v,t)] + D_{1} \frac{\partial^2 P(v,t)}{\partial v^2}-
\beta D_{2} v \frac{\partial^3 P(v,t)}{\partial v^3} \hspace{0.2cm},
\end{equation}

\noindent
where
\begin{equation}
\left. \begin{array}{ccc}
D_{1} & = & \alpha^2 ( c_{12}-3c_{3}\beta)\\
D_{2} & = & \alpha^2 c_{3}
\end{array} \right \} \hspace{0.2cm}.
\end{equation}

We now transform the Eq.(46) to Fourier space by defining the conditional 
probability $P(v,t|v_{0},0)$ and its Fourier transform as
\begin{equation}
P(v,t|v_{0},0) = \frac{1}{2\pi} \int_{-\infty}^{+\infty} dk e^{ikv}
{\tilde P}(k,t|v_{0},0) \hspace{0.2cm},
\end{equation}

\noindent
to obtain
\begin{equation}
\frac{\partial}{\partial t} {\tilde P}(k,t|v_{0},0) = -\beta (k+ D_{2} k^3)
\frac{\partial {\tilde P}}{\partial k} - (D_{1} +3\beta D_{2}) k^2 {\tilde P} \hspace{0.2cm}.
\end{equation}

The linear partial differential equation (49) can be solved by the method of 
characteristics. For the initial condition (at time $t=0$)
\begin{equation}
P(v,0|v_{0},0)=\delta (v-v_{0})\hspace{0.2cm},
\end{equation}

\noindent
the solution is 
\begin{equation}
{\tilde P} (k,t|v_{0},0) = \frac{1}{(1+Bk^2)^A} \exp\left [ -ikv_{0}
\sqrt{\frac{f(t)}{1+Bk^2}} \; \right ] \hspace{0.2cm},
\end{equation}

\noindent
where
\begin{equation}
\left. \begin{array}{cccc}
& f(t) & = & e^{-2\beta t}\\

& A & = & c_{12}/2 \beta c_{3}\\
{\rm and} & B & = & \alpha^2 c_{3} \{ 1- f(t) \}
\end{array} \right \} \hspace{0.2cm}.
\end{equation}

It is easy to check that Eq.(51) satisfies 
\begin{eqnarray*}
{\tilde{P}}^\ast (k,t|v_0,0) = {\tilde{P}} (-k,t|v_0,0)
\end{eqnarray*}

and the characteristic function (51) is of stable form.

The conditional probability density $P(v,t|v_{0},0)$ is obtained by inverse
Fourier transformation of Eq.(51) and is given by,
\begin{equation}
P(v,t|v_{0},0) = \frac{1}{2\pi} \int_{-\infty}^{+\infty} dk
\frac{1}{(1+Bk^2)^A} 
\exp \left [ \; ikv - ikv_{0}
\sqrt{\frac{f(t)}{1+Bk^2}}  \; \right ] \hspace{0.2cm}.
\end{equation}

It is evident that $P(v,t|v_{0},0)$ is a Fourier transform of a 
stable characteristic function. Hence the solution (53) forms a stable 
distribution in the variable $v$. The Eq.(53) is one of the important results 
of this paper.

Although an explicit expression for $P(v,t|v_{0},0)$ is difficult to obtain,
closed-form solutions for $P(v,t|v_{0},0)$ for stationary state can be easily 
obtained. In the long time limit the characteristic function (51) reduces to 
its asymptotic form
\begin{equation}
{\tilde P} (k,\infty) = \frac{1}{(1+ D_{2} k^2)^A} \hspace{0.2cm},
\end{equation}

\noindent
which results a steady state distribution of stable form. Explicitly for small 
$A$, i.e., large $\beta$ this is given by
\begin{equation}
P_{\rm st}(v) = \frac{ |v|^{A+1} }
{ 2^{A} \hspace{0.1cm} D_{2}^{\frac{A}{2}} \hspace{0.1cm} \Gamma(A) 
\hspace{0.1cm} v^2 } \hspace{0.1cm}e^{-\frac{|v|}{\sqrt{D_{2}}}} \hspace{0.2cm}.
\end{equation}

It is interesting to note that the dominant behavior of $P_{\rm st}(v)$ 
for small $v$. This is given by a power law of the form
\begin{equation}
\left. \begin{array}{ccc}
P_{\rm st} (v) & \sim & \frac{ |v|^{A+1}}{v^2}\\
& \sim & |v|^{-1+A}
\end{array} \right \} \hspace{0.2cm}.
\end{equation}

Such power law behavior is also apparent for Levy processes [9,11] but 
for the large $v$ regime.

Although explicit solution for probability distribution $P(v,t|v_{0},0)$
is difficult to obtain for arbitrary time, however, a few statistical 
properties of 
the process can be obtained from the calculation of variances and higher 
moments. For convenience, we define such moments by subtracting the mean 
motion of the variables, i.e., we calculate the moments of $\Delta v (=
v - v_{0} e^{-\beta t})$. Thus we write
\begin{equation}
\langle |\Delta v|^{m}\rangle = \int_{-\infty}^{+\infty}
(v-v_{0} e^{-\beta t})^{m} P(v,t|v_{0},0) \hspace{0.2cm},
\end{equation}

\noindent
or more explicitly
\begin{eqnarray}
\langle |\Delta v|^{m}\rangle = \frac{1}{2\pi} \int_{-\infty}^{+\infty}  dk 
\frac{1}{(1+Bk^2)^{A}}
\exp \left [ ikv_{0} e^{-\beta t} \left\{ 1- \left (1+Bk^2 \right )^{-\frac
{1}{2}}\right \} \right ] \nonumber\\
\int_{-\infty}^{+\infty} d \Delta v |\Delta v|^{m} e^{ik \Delta v} \hspace{0.2cm}.
\end{eqnarray}

After some algebra we get
\begin{equation}
\langle |\Delta v|^{m}\rangle = (-i)^m \int_{-\infty}^{+\infty}  dk 
\frac{1}{(1+Bk^2)^{A}}
\exp \left [ ikv_{0} e^{-\beta t} \left\{ 1- \left (1+Bk^2 \right )^{-\frac
{1}{2}}\right \} \right ] \frac{\partial ^m \delta (k)} {\partial k^m} .
\end{equation}

In principle, using the property of Dirac $\delta$-function and appropriate 
integrations any moment can be calculated from the above relation. We quote 
the results explicitly for the first three moments.
\begin{equation}
\left. \begin{array}{ccccc}
{\rm For} & m=1 & \langle |\Delta v| \rangle & = & 0\\
{\rm For} & m=2 & \langle |\Delta v|^{2}\rangle & = &  \frac{\alpha^2 c_{12}} 
{\beta} ( 1-e^{-2 \beta t} )\\
{\rm For} & m=3 & \langle |\Delta v|^{3}\rangle  & = & 3\alpha^2 c_{3} v_{0}
e^{-\beta t} ( 1-e^{-2 \beta t} ) \end{array} \right \} \hspace{0.2cm}.
\end{equation}

It is thus evident that unlike Levy processes [9,11] the moments are finite.

We thus observe that because of the linear dissipation $\beta$, a system driven
by adiabatic noise reaches a steady state which is stable. However, since the
noise is of external origin, the outward flow of energy due to linear 
dissipation cannot balance the inward flow of energy supplied by the adiabatic 
fluctuations and hence a fluctuation-dissipation relation cannot be conceived 
in this case.

\vspace{0.5cm}


\begin{center}
\large{
\bf {VI.\hspace{0.2cm} Conclusions}}
\end{center}

\vspace{0.5cm}

In conclusion, we consider herein a dynamical system driven by external 
adiabatic fluctuations. Based on the `adiabatic following approximation' 
we have made a systematic separation of time-scales to carry out an 
expansion in $\alpha |\mu|^{-1}$ to obtain a linear differential equation
for the average solution, where $\alpha$ is the strength of 
fluctuation and $|\mu|$ is the largest eigenvalue of the unperturbed system. 
The main results of this study can be summarized as follows :

(i) The probability distribution functions obey the differential equations of
motion which contain third-order terms beyond the usual Fokker-Planck
terms. The adiabatic fluctuations thus may give rise to non-Gaussian noise.

(ii) We have examined in detail the corresponding equation in velocity space
and the characteristic function is shown to obey a simple third-order differential
equation which can be solved exactly in closed form. The characteristic 
function is found to be of stable form. 

(iii) Although third-order noise, in general, leads to serious interpretative
difficulties in terms of truly probabilistic considerations in several
cases, we show that in the present problem of adiabatic stochasticity in 
velocity space, statistical properties of the processes are more transparent.
It is of special interest to note that in contrast to Levy process all the
variances and higher moments are finite and probability distribution is of 
stable form.

(iv) Because of linear dissipation, the system driven by adiabatic fluctuations
reaches a stable steady state. 

(v) For small arguments the probability distribution function obeys a power 
law behavior which is reminiscent of Levy processes.

The stochastification by adding rapid fluctuating terms had been applied
earlier to a wide variety of physical problems described by linear relaxation
equations [15], hydrodynamic equations [16], Maxwell equations in a 
medium [17], 
Boltzmann equation [18] etc. Our present analysis shows that the 
present method might reveal interesting consequences in such cases
where the added fluctuating terms, in question, are adiabatically slow. We
hope to address such issues in future communications.

{\bf Acknowledgments} : Partial financial support from the Department of 
Science and Technology, Government of India is thankfully acknowledged. One of
us (S.K.Banik) is thankful to Prof. J. K. Bhattacharjee (Dept. of
Theoretical Physics) for helpful discussions.

\newpage

\begin{center}
\large{\bf References}
\end{center}

\vspace{0.5cm}

\begin{enumerate}
\item Selected Papers on Noise and Stochastic Processes, edited by N. Wax
(Dover, New York, 1954).
\item Stochastic Processes in Physics and Chemistry, N. G. van Kampen
(North-Holland, Amsterdam, 1992).
\item R. C. Bourret, U. Frisch and A. Pouquet, Physica {\bf 65}, 303 (1973) ;
O. J. Heilman and N. G. van Kampen, Physica {\bf 77}, 279 (1974) ; M. R. 
Cruty and K. C. So, Phys. Fluids. {\bf 16}, 1765 (1973) ; A. Brissaud and U.
Frisch, J. Math. Phys. {\bf 15}, 524 (1974) ; H. F. Arnoldus and G. Nienhus,
J. Phys. B {\bf 16}, 2325 (1983) ; J. Phys. A {\bf 19}, 1629 (1986) ; M. 
Rahman, Phys. Rev. {\bf E52}, 2486 (1995) ; R. Walser, H. Ritsch, P. Zoller
and J. Cooper, Phys. Rev. {\bf A45}, 468 (1992) ; V. Berdichevsky and M.
Gitterman, Europhysics Letts. {\bf 36}, 161 (1996)
\item M. D. Crisp, Phys. Rev. {\bf A8}, 2128 (1973).
\item S. K. Banik and D. S. Ray, J. Phys. A : Math. and General, {\bf 31},
3937 (1998).
\item R. F. Pawula, Phys. Rev. {\bf 162}, 186 (1967).
\item See, for example, C. W. Gardiner, Handbook of Stochastic Methods,
p.299, (Springer-Verlag, Berlin, 1983).
\item G. Ryskin, Phys. Rev. {\bf E56}, 5123 (1997) ;
N. G. van Kampen and I. Oppenheim, J. Math. Phys. {\bf 13}, 842 (1972).
\item B. J. West and V. Seshadri, Physica {\bf 113A}, 203 (1982).
\item R. C. Bourret, Can. J. Phys. {\bf 40}, 782 (1962) ; Nuovo. Cim. {\bf 26},
1 (1962).
\item Levy Flights and Related Topics in Physics, edited by M. F. Shlesinger,
G. M. Zaslavsky and U. Frisch (Springer, New York, 1995) ;
B. J. West and W. Deering, Phys. Rep. {\bf 246}, 1 (1994).
\item N. G. van Kampen, Phys. Rep. {\bf 24}, 171 (1976).
\item H. A. Kramers, Physica {\bf 7}, 284 (1940).
\item M. Bixon and R. Zwanzig, J. Stat. Phys. {\bf 3}, 245 (1971).
\item L. Onsager and S. Machlup, Phys. Rev. {\bf 91}, 1505 (1953) ;
{\bf 91}, 1512 (1953).
\item Fluid Mechanics, L. D. Landau and E. M. Lifshitz, (Pergamon, Oxford, 
1959) Ch. 17 .
\item Electrodynamics, L. D. Landau and E. M. Lifshitz, (Pergamon, Oxford, 
1960) Ch. 13 .
\item R. F. Fox and G. E. Uhlenback, Phys. Fluids. {\bf 13}, 2881 (1970).
\item D. Grischkowsky, Phys. Rev. Letts. {\bf 24}, 866 (1970); 
D. Grischkowsky, E. Courtens and J. Armstrong, Phys. Rev. Letts. {\bf 31},
422 (1973).
\end{enumerate}

\end{document}